\documentclass{ws-procs975x65}

\begin{document}



\title{A teleparallel representation of the Weyl Lagrangian}

\author{DMITRI VASSILIEV}

\address{Dept of Mathematics,
University College London,
Gower Street,
London WC1E 6BT,
UK\\
\email{D.Vassiliev@ucl.ac.uk}}


\begin{abstract}
The main result of the paper is a new representation of the Weyl
Lagrangian (massless Dirac Lagrangian). As the dynamical variable we use
the coframe, i.e. an orthonormal tetrad of covector fields.
We write down a simple Lagrangian -- wedge product of axial torsion with a
lightlike element of the coframe -- and show that variation of the
resulting action with respect to the coframe
produces the Weyl equation. The advantage of our approach
is that it does not require the use of spinors, Pauli matrices or
covariant differentiation. The only geometric concepts we use are
those of a metric, differential form, wedge product and exterior derivative.
Our result assigns a variational meaning to the tetrad
representation of the Weyl equation suggested by J.~B.~Griffiths and R.~A.~Newing.
\end{abstract}

\bodymatter

\section{Traditional model for the neutrino}

Throughout this paper we work on a 4-manifold $M$ equipped with
prescribed Lorentzian metric $g$.

The accepted mathematical model
for a neutrino field is the following linear partial differential
equation on $M$ know as the \emph{Weyl equation}:
\begin{equation}
\label{Weyl's equation}
i\sigma^\alpha{}_{a\dot b}\{\nabla\}_\alpha\xi^a=0.
\end{equation}
The corresponding Lagrangian is
\begin{equation}
\label{Weyl's Lagrangian}
L_\mathrm{Weyl}(\xi):=
\frac i2
(\bar\xi^{\dot b}\sigma^\alpha{}_{a\dot b}\{\nabla\}_\alpha\xi^a
-
\xi^a\sigma^\alpha{}_{a\dot b}\{\nabla\}_\alpha\bar\xi^{\dot b})
*1.
\end{equation}
Here
$\sigma^\alpha$, $\alpha=0,1,2,3$, are Pauli matrices,
$\xi$ is the unknown spinor field,
and $\{\nabla\}$ is the covariant derivative with
respect to the Levi-Civita connection:
$\{\nabla\}_\alpha\xi^a:=
\partial_\alpha\xi^a
+\frac14\sigma_\beta{}^{a\dot c}
(\partial_\alpha\sigma^\beta{}_{b\dot c}
+\{\Gamma\}^\beta{}_{\alpha\gamma}\sigma^\gamma{}_{b\dot c})\xi^b$
where $\{\Gamma\}^\beta{}_{\alpha\gamma}$ are Christoffel symbols
uniquely determined by the metric.

\section{Teleparallel model for the neutrino}

The purpose of our paper is to give an alternative representation
of the Weyl equation (\ref{Weyl's equation})
and the Weyl Lagrangian (\ref{Weyl's Lagrangian}).
To this end, we follow \cite{MR0332092} in
introducing instead of the spinor field a different unknown
-- the so-called \emph{coframe}. A coframe is a quartet of real
covector fields
$\vartheta^j$, $j=0,1,2,3$,
satisfying the constraint
\begin{equation}
\label{constraint for coframe}
g=o_{jk}\,\vartheta^j\otimes\vartheta^k
\end{equation}
where $o_{jk}=o^{jk}:=\mathrm{diag}(1,-1,-1,-1)$.
In other words, the coframe is a field of orthonormal bases with
orthonormality understood in the Lorentzian sense.

We define an affine connection and corresponding covariant
derivative $|\nabla|$ from the conditions
\begin{equation}
\label{defining condition for connection}
|\nabla|\vartheta^j=0.
\end{equation}
Let us emphasize that we
employ holonomic coordinates, so in explicit form
conditions (\ref{defining condition for connection}) read
$\partial_\alpha\vartheta^j_\beta-|\Gamma|^\gamma{}_{\alpha\beta}\vartheta^j_\gamma=0$
giving a system of linear algebraic equations for the unknown
connection coefficients $|\Gamma|^\gamma{}_{\alpha\beta}$.
The connection defined by
the system of equations~(\ref{defining condition for connection})
is called the \emph{teleparallel} or \emph{Weitzenb\"ock} connection.

Let $l$ be a nonvanishing real lightlike teleparallel covector field
($l\cdot l=0$, $|\nabla|l=0$).
Such a covector field can be written down explicitly as $l=l_j\vartheta^j$
where the $l_j$ are real constants (components of the covector
$l$ in the basis $\vartheta^j$),
not all zero, satisfying
\begin{equation}
\label{constraint for constants}
o^{jk}l_jl_k=0.
\end{equation}
We define our Lagrangian as
\begin{equation}
\label{Lagrangian as function of vartheta and c}
L(\vartheta^j,l_j)=l_io_{jk}\,
\vartheta^i\wedge\vartheta^j\wedge d\vartheta^k
\end{equation}
where $d$ stands for the exterior derivative.
Note that
$\frac13\,o_{jk}\,\vartheta^j\wedge d\vartheta^k$
is the axial (totally antisymmetric) piece of torsion
of the teleparallel connection.
Let us emphasize that formula
(\ref{Lagrangian as function of vartheta and c})
does not explicitly involve connections or covariant derivatives.

The Lagrangian (\ref{Lagrangian as function of vartheta and c})
is a rank 4 covariant
antisymmetric tensor so it can be viewed as a 4-form and
integrated over the manifold
$M$ to give the action
$S(\vartheta^j,l_j):=\int L(\vartheta^j,l_j)$.
Variation with respect to
the coframe $\vartheta^j$
subject to the constraint (\ref{constraint for coframe})
produces an
Euler--Lagrange equation which we write symbolically as
\begin{equation}
\label{field equation 1}
\partial S(\vartheta^j,l_j)/\partial\vartheta^j=0.
\end{equation}
The explicit form of the field equation
(\ref{field equation 1}) is given in Griffiths' and Newing's paper \cite{MR0332092}.

\section{Equivalence of the two models}

Let us define the spinor field $\xi$ as the solution of the system
of equations
\[
|\nabla|\xi=0,
\qquad
\sigma_{\alpha a\dot b}\xi^a\bar\xi^{\dot b}=\pm l_\alpha
=\pm l_j\vartheta^j_\alpha
\]
where
$|\nabla|_\alpha\xi^a:=
\partial_\alpha\xi^a
+\frac14\sigma_\beta{}^{a\dot c}
(\partial_\alpha\sigma^\beta{}_{b\dot c}
+|\Gamma|^\beta{}_{\alpha\gamma}\sigma^\gamma{}_{b\dot c})\xi^b$.
The above system
determines the spinor field $\xi$ uniquely
up to a complex constant factor of modulus $1$.
This non-uniqueness is acceptable because we will be substituting
$\xi$ into the Weyl equation~(\ref{Weyl's equation})
and Weyl Lagrangian
(\ref{Weyl's Lagrangian})
which are both $\mathrm{U}(1)$-invariant.
We will call $\xi$ the spinor field \emph{associated}
with the coframe $\vartheta^j$.

The main result of our paper is the following

\begin{theorem}
\label{main result}
For any coframe $\vartheta^j$ we have
$
L(\vartheta^j,l_j)=\pm4L_\mathrm{Weyl}(\xi)
$
where $\xi$ is the associated spinor field.
The coframe satisfies the field equation
(\ref{field equation 1})
if and only if the associated spinor field satisfies
the Weyl equation (\ref{Weyl's equation}).
\end{theorem}

The proof \cite{vassilievPRD} of Theorem~\ref{main result} is based
on the observation that our Lagrangian (\ref{Lagrangian as function of vartheta and c}) is
invariant under the action of a certain class of local (i.e. with
variable coefficients) transformations of the coframe, the class in
question being the subgroup $B^2$ of the Lorentz group
\cite{MR867684}. This means that coframes come in equivalence
classes (cosets) and the nature of these cosets is such that they
can be identified with spinors.

\section{Discussion}

Our Lagrangian (\ref{Lagrangian as function of vartheta and c}) has
the unusual feature that it depends on a quartet of real parameters
$l_j$ which can be chosen arbitrarily as long as they satisfy the
condition~(\ref{constraint for constants}).
This parameter dependence requires an explanation.
One possible explanation of the physical nature of the $l_j$'s
is sketched out below.

Consider the Lagrangian
\begin{equation}
\label{axial torsion squared}
L(\vartheta^j):=\|o_{jk}\,\vartheta^j\wedge d\vartheta^k\|^2*1
\end{equation}
(``axial torsion squared'').
Putting $S(\vartheta^j):=\int L(\vartheta^j)$
and varying with respect to the coframe $\vartheta^j$
subject to the constraint (\ref{constraint for coframe})
we get an Euler--Lagrange equation which we write symbolically as
\begin{equation}
\label{field equation for quadratic Lagrangian}
\partial S(\vartheta^j)/\partial\vartheta^j=0.
\end{equation}

It turns out that in
the case of \emph{Minkowski} metric one can construct an explicit
solution of (\ref{field equation for quadratic Lagrangian})
as follows. Let $\bm{\vartheta}^j$ be a constant reference coframe
and let $l\ne0$ be a constant real lightlike covector;
here ``constant'' means ``parallel with respect to the
Levi-Civita connection induced by the Minkowski metric''.
Of course,
\begin{equation}
\label{origin of the l_j's}
l=l_j\bm{\vartheta}^j
\end{equation}
for some real constants $l_j$
satisfying the condition (\ref{constraint for constants}).
Perform a rigid Lorentz transformation
$\bm{\vartheta}^j\mapsto\tilde{\bm{\vartheta}}{}^j=\Lambda^j{}_k\bm{\vartheta}^k$
so that $l=\omega(\tilde{\bm{\vartheta}}{}^0+\tilde{\bm{\vartheta}}{}^3)$
for some $\omega\ne0$, put
\[
\begin{pmatrix}
\tilde\vartheta{}^0\\
\tilde\vartheta{}^1\\
\tilde\vartheta{}^2\\
\tilde\vartheta{}^3
\end{pmatrix}:=
\begin{pmatrix}
{\ 1\ }&0&0&{\ 0\ }\\
0&{\ \cos\omega(x^0+x^3)\ }&{\ \sin\omega(x^0+x^3)\ }&0\\
0&{\ -\sin\omega(x^0+x^3)\ }&{\ \cos\omega(x^0+x^3)\ }&0\\
{\ 0\ }&0&0&{\ 1\ }
\end{pmatrix}
\begin{pmatrix}
\tilde{\bm{\vartheta}}{}^0\\
\tilde{\bm{\vartheta}}{}^1\\
\tilde{\bm{\vartheta}}{}^2\\
\tilde{\bm{\vartheta}}{}^3
\end{pmatrix}
\]
where $\ x^j:=\int\tilde{\bm{\vartheta}}{}^j\cdot dy\ $
(here $y^\alpha$ are arbitrary local coordinates on the manifold $M$ and $x^j$
are Minkowskian coordinates on $M$ associated with the constant
coframe~$\tilde{\bm{\vartheta}}{}^j$),
and, finally, set
$\vartheta^j:=(\Lambda^{-1})^j{}_k\tilde\vartheta{}^k$.
Straightforward calculations show that the coframe $\vartheta^j$
is indeed a solution of (\ref{field equation for quadratic Lagrangian}).
We call this solution \emph{plane wave}
with \emph{momentum}~$l$. Note that for a plane wave
$\ o_{jk}\,\vartheta^j\wedge d\vartheta^k=\pm2*l\ $
and $\ l=l_j\vartheta^j\ $ where the $l_j$ are the original constants from
(\ref{origin of the l_j's}).

Now consider a perturbation of a plane wave. This perturbation can be the result of
either a) us looking for a wider class of solutions or b) the metric
ceasing to be Minkowski. Application of a formal perturbation argument
to the Lagrangian~(\ref{axial torsion squared}) with
$\ o_{jk}\,\vartheta^j\wedge d\vartheta^k\mp2l_j*\vartheta^j\ $
as small parameter
gives the Lagrangian (\ref{Lagrangian as function of vartheta and c}).

\vfill


\begin{thebibliography}{00}

\bibitem{MR0332092}
J. B. Griffiths and R. A. Newing,
{\it J. Phys. A\/}: {\it Gen. Phys.} {\bf 3}, 269 (1970).

\bibitem{vassilievPRD} D. Vassiliev,
{\it Phys. Rev.} {\bf D75}, 025006 (2007); gr-qc/0604011.

\bibitem{MR867684}
A. L. Besse,
{\it Einstein manifolds}
(Springer, 1987).
\end{thebibliography}
\end{document}